\documentclass[a4paper,twocolumn]{esapub2005} 
\usepackage{graphics,epsfig,lscape,times}
\usepackage{natbib}

\def\deg   {$^o$\ }

\title{Multiwavelength appearance of Vela Jr.: Is it up to expectations?}
\author{Anatoli F. Iyudin}
\affil{Skobeltsyn Institute of Nuclear Physics, Moscow State University, Vorob'evy Gory, 119992 Moscow, Russian Federation}
\author{B. Aschenbach}
\author{V. Burwitz}
\author{K. Dennerl}
\author{M. Freyberg}
\author{F. Haberl}
\affil{Max-Planck-Institut f\"ur Extraterrestrische Physik, Postfach 1312, D-85741 Garching, Germany}
\author{M. Filipovic}
\affil{CSIRO Australia Telescope National Facility, PO Box 76, Epping NSW1710, Australia; University of Western Sydney, Locked Bag 1797, Penrith South, DC, NSW 1797, Australia}

\begin{document}

\keywords{galactic supernova remnants; radio, optical, and X-ray emission; nucleosynthesis; $\gamma$-ray lines}

\maketitle

\begin{abstract}
Vela Jr. is one of the youngest and likely nearest among the known galactic supernova remnants (SNRs). 
Discovered in 1997 it has been studied since then at quite a few wavelengths, that spread over almost 20 decades in energy.
\par
Here we present and discuss Vela Jr. properties revealed by these multiwavelength observations, and confront them with the SNR model expectations. 
\par
Questions that remained unanswered at the time of publication of the paper by Iyudin et al. [25], e.g. what is the nature of the SNR's proposed
central compact source CXOU J085201.4-461753, and why is the ISM absorption column density apparently associated with RX J0852.0-4622 much
greater than the typical column of the Vela SNR, can be addressed using latest radio and X-ray observations of Vela Jr.. These, and other related questions are addressed in the following.
\end{abstract}

\section{Introduction}
RX J0852.0-4622, 
referred to as G266.2-1.2 in Green's SNR catalogue (http://www.mrao.cam.ac.uk/surveys/snrs/), and occasionally as Vela Jr., located at the 
south-eastern corner of the Vela SNR [2] was one of the discoveries of 
the {\it{ROSAT}} all-sky survey. 
It is also the second brightest excess in the $\sim$6 years all-sky map of {\it{COMPTEL}} in the 1.157 MeV  line emission, called 
GRO J0852-4642, that coincides with 
RX J0852.0-4622 [19, 20]. Since the 1.157 MeV line of $^{44}$Ti is exclusively produced in supernovae it is very likely 
that RX J0852.0-4622 and GRO J0852-4642 are the same object which was created in one supernova explosion. 
The combined analysis of the X-ray data and the $\gamma$-ray data led to the suggestion that RX J0852.0-4622 could be 
the remnant of the nearest supernova in recent history [3] with a best estimate for the distance of 
200 pc and an age of 680 years. 
\par
The detection of 
$^{44}$Ti in Cas A [17, 18] has been supported by the {\it{Beppo-SAX}} measurements of the $\sim$68
and $\sim$78 keV X-ray lines [50] which are produced in the first decay of $^{44}$Ti in the decay chain $^{44}$Ti$\rightarrow$$^{44}$Sc$\rightarrow$$^{44}$Ca, and recently by IBIS/ISGRI [38, 51]. The detection of the 1.157
MeV $\gamma$-ray line from Cas A was the first discovery of $^{44}$Ti in a
young galactic SNR, and as such it provides an
essential calibration of nucleosynthesis model calculations. 
With the discovery of RX J0852.0-4622/GRO J0852-4642 we may have a second example, which is still needs the confirmation by independent measurements. SN1987A might be the third case of strong $^{44}$Ti line emission, which can be probed in the future by more sensitive $\gamma$-ray instruments.
\par
Multiple observations of Vela Jr. in the radio band confirmed the shell-like structure and supported the identification 
of RX J0852.0-4622 as a SNR [8,11,12,13,14,16,46]. Furthermore a good correlation 
between X-ray and radio brightness was found (c.f. Fig. 1). 

\begin{figure}
  \centering
\includegraphics[bb=70 370 532 745,width=8.5cm,clip]{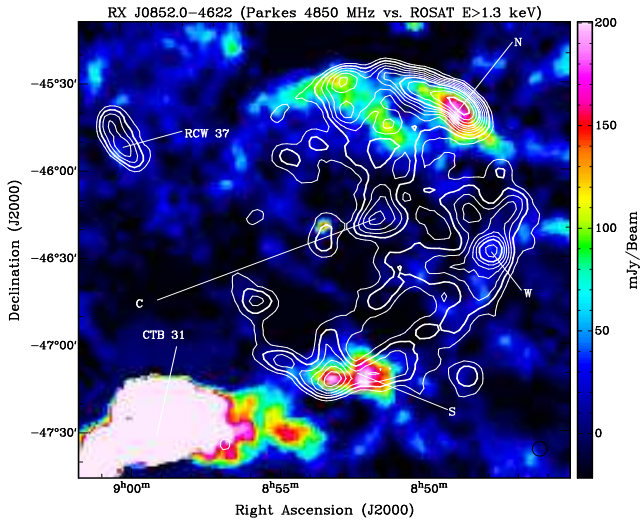}
\caption{The Parkes image (colour) of RX J0852.0-4622 overlaid with ROSAT PSPC contours (E$>$1.3 keV). 
The synthesized beam of the Parkes observations is 5$'$ (lower left corner) with an r.m.s noise (1 $\sigma$)
of $~$10 mJy. X-ray contours (white) are 0.8, 1, 1.2, 1.4, 1.6, 1.8, 2, 2.3, 2.6 and 3 in units of 10$^{-4}$
ROSAT PSPC counts s$^{-1}$ arcmin$^{-2}$.}
\end{figure}
\par
X-ray spectra of this SNR taken with {\it{ROSAT}}, ASCA, {\it Chandra} and XMM-Newton are well described by a power law 
[3,4,22,33,44,48]. 
The best-fit with a single power law results in an interstellar absorption column density of 3.9$\times$10$^{21}$ cm$^{-2}$ which is at least a factor of six higher 
than the highest value of $\sim$6$\times$10$^{20}$ cm$^{-2}$ observed anywhere else in the Vela SNR [31] for the thermal emission fit. 
\par
Based on the {\it{ROSAT}} data
Aschenbach [2] suggested the presence of a central point source, which later was confirmed
by {\it{Chandra}} measurements [34]. The spectrum of the proposed compact central source (CCS) CXOU J085201.4-461753 suggests a neutron star as the emitter [23,5,6], which is supported 
by the absence of any optical counterpart brighter than R$\sim$26 [30].
If this object is the compact remnant of the supernova which created RX J0852.0-4622 the supernova was of the 
core-collapse type. Like for the north-western rim the absorbing column density of (3.45$\pm$0.15)$\times$10$^{21}$ cm$^{-2}$) inferred from the spectrum of CXOU J085201.4-461753 is significantly higher
by a factor of about six [23] than the column densities of $\sim$6$\times$10$^{20}$ cm$^{-2}$ measured for the Vela SNR by Lu \& Aschenbach  [29],
which suggests that at least CXOU J085201.4-461753 and possibly RX J0852.0-4622 are at a greater distance than
the Vela SNR, which is located at a distance of 390$\pm$100 pc, as was recomended by Cha \& Sembach [7] based on observations of 60 O stars.
Pozzo et al. [36] concluded that the distance to RX J0852.0-4622 is $\approx$430$\pm$60 pc, which is consistent with
the distance to the Vela OB2 association that was derived to 420$\pm$30 pc by Woermann et al. [52], or with the distance to Trumpler 10 (Tr 10) OB association that was derived by de Zeeuw et al. [56]. 
\par
On the other hand, Redman et al. [37] suggested, as a possibility, that the optical nebula RCW 37 was generated
by the blast wave of RX J0852.0-4622 impacting the shell of the Vela SNR, i.e. RX J0852.0-4622 is embedded in the Vela SNR. Moreover, the X-ray spectrum of rather bright, nebular region C (Fig. 1), which is slightly eastward of CXOU J085201.4-461753 (Fig. 2), has an appearance typical of the Vela SNR shrapnel spectrum [25, 26], but has an absorption column of 
$\sim$5.0$\times$10$^{21}$ cm$^{-2}$ (Iyudin et al. 2006, in preparation), that is even higher than the column to the CCS. This fact, among others, suggests a high degree of clumpiness of the ISM towards Vela Jr. along different line of sights in this complex region containing a number of molecular, HI and CS clouds [10,31,55,57]. We note, that this high absorption column is derived for the line sight very close to the line sight towards CCS. If the column of $\sim$5.0$\times$10$^{21}$ cm$^{-2}$ corresponds to the thikness of two ejecta shells, or of two walls of the wind bubble, an absorption column towards the CCS of around $\sim$2.5$\times$10$^{21}$ cm$^{-2}$ is expected which is not far from the columns measured for the CCS [6].
\par
Concerning the age, there has been the suggestion, that the supernova which led to RX J0852.0-4622 is responsible 
for a previously unidentified spike in nitrate concentration measured in an Antarctic ice core. The precipitation 
occurred around the year 1320. Other nitrate spikes could be associated with historical supernovae.  
One of the issues in this context is of course the detection of the 1.157 MeV line, because together with the
$^{44}$Ti yield it dominates the estimate of the age.
\par
To get more insight into the questions of distance, age and progenitor of this SNR observations of the X-ray bright rims and of the central region were carried out by {\it XMM-Newton} [5,6] and by {\it Chandra} [4,23,33,34]. 

\begin{figure}[hbt]
  \centering
\includegraphics[bb=20 20 575 564,width=8.0cm,clip]{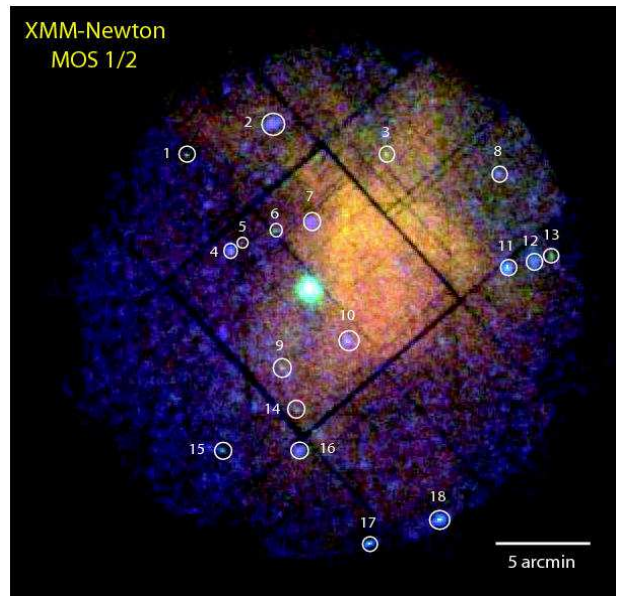}
\caption{XMM-Newton MOS1/2 false color image of the inner 30 arcmin central region of RX J0852.0-4622 (red: 0.3-0.75 keV; green: 0.75-2 keV, and blue: 2-10 keV). The central bright source is CXOU J085201.4-461753. Other 18 point-like sources, marked by circles, are detected in the field of view. The angular binning factor of image is 4 arcsec [6].
}
\end{figure}

\section{Radio observations}  
Radio observations of RX J0852.0-4622 starting with the publication of Dubner et al. [10], and on the Vela and Gum Nebula regions by Yamaguchi et al. [55], Moriguchi et al. [31] shed light on the SNR immediate environment, demonstrating the presence of a multitude of different clouds in the ISM surrounding the SNR. Especially noteworthy the paper by Dubner et al. [10] that shows HI clouds of an angular size comparable to that of RX J0852.0-4622 whose boundary is likely to be interacting with the RX J0852.0-4622 blast wave (see Fig. 3). The presence of many point-like X-ray sources in the vicinity of the forward shock can be used to demonstrate the increase of the absorbing column along the radii going from the center of the SNR to the point source No. 1 near the NW rim. This fact clearly supports the idea of the forward shock just beginning to interact with the outer cloud, presumably the HI cloud discovered by Dubner et al. [10].
\par
Another interesting recent radio result was produced by Reynoso et al. [42] who presented new, more sensistive observations of the RX J0852.0-4622 central region at a few radio frequencies. 
The major result of their study is that no central radio nebulosity can be found at or near the position of the H$_{\alpha}$ nebulosity of Pellizzoni et al. [35]. Instead, a nebulosity was found that is tentatively identified as a ``butterfly''-type planetary nebula positioned quite nearby of Wray 16-30, and CXOU J085201.4-461753. Implications of this new discovery are not completely clear at the moment, but previous claim of the PWN type radio nebulosity has been withdrawn [42]. 

\section{X-ray observations}  

The diameter of RX J0852.0-4622 is $\sim$2\deg which is significantly larger than the $\sim 30'$ field of view of the EPIC instruments 
on board of {\it{XMM-Newton}}. Four different pointings on the brightest sections of the remnant were carried out in the GTO program, three 
of which were directed to the rim, i.e. the northwest (NW), the west (W) and the south (S) and the fourth pointing was towards the 
center (C) (Fig. 2). 
\begin{table*}[hbt]
\caption{{\it{XMM-Newton}} XMM GTO observations of RX J0852.0-4622}

\smallskip
\begin{center}
\begin{tabular*}{71mm}{||c|cc|c||@{\extracolsep\fill}} \hline
  &\multicolumn{2}{|c|} {Pointing (J2000)}  & Expos. \\
\cline{2-4}
  \raisebox{1.5ex}[0pt]{Rim} & RA & DEC & ksec   \\
\hline
\hline
NW & 08h48m58s  & -45d39m03s  & 31.76   \\
\hline
West& 08h47m45s  & -46d28m51s  & 35.96   \\
\hline
South & 08h53m14s  & -47d13m53s  & 47.13  \\
\hline
Center& 08h51m50s  & -46d18m45s  & 24.11   \\
\hline
\hline
\end{tabular*}
\end{center}
\end{table*}

The observations were carried out between April 24 and April 27, 2001. The EPIC-PN camera [45] was operated in 
extended full frame mode and the medium filter was in place. The EPIC-MOS1 and -MOS2 cameras [49] were used in full frame mode with the medium 
filter as well. Further details of the observations are given in Table 1. These observations taken during XMM-Newton revolutions 0252, and 0253, were augmented by observations of the NW rim to calibrate the low energy response of the EPIC PN. We have used some fraction of these observations to analyse the spectra and the radial profile of the NW rim. 
\begin{figure}[hbt]
  \centering
\includegraphics[bb=0 0 810 567,width=8.0cm,angle=0,clip]{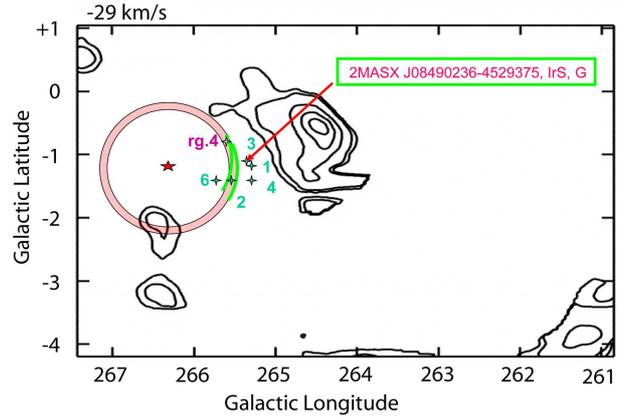}
\caption{Relative position of the HI cloud near RX J0852.0-4622 found by Dubner et al. (1998), relative to the SNR as a whole, relative to the NW rim of SNR, and to some previously unknown and yet unidentified point-like X-ray sources (Iyudin et al. 2006, in preparation). Only one source (3) has tentatively been identified with the galaxy 2MASX J08490236-4529375.
}
\end{figure}

\subsection{X-ray spectral fits}

RX J0852.0-4622 is located in the south-eastern corner of the Vela SNR, which actually completely covers RX J0852.0-4622. At low energies 
the X-ray surface brightness of the Vela SNR is much higher than that of RX J0852.0-4622, so that RX J0852.0-4622 becomes visible only 
above $\sim$1 keV in the {\it{ROSAT}} images. This is a major complication for any spectral fit to RX J0852.0-4622. We have followed three options. 
In a first approach we chose fields for the background to be subtracted  which are definitely outside the area covered by RX J0852.0-4622 but inside 
the Vela SNR. The corresponding fits did not converge to a unique solution, when different fields for the background were chosen. This is likely 
to be due to the Vela spectra changing on a scale of a few arcminutes [29]. In a second approach we fit the spectra of RX J0852.0-4622 with a three component model. This model consists of one thermal component with an associated 
interstellar absorption column density which together represent the low energy Vela SNR emission; the second component is again thermal emission with an added 
power law (third component) to represent the higher energy emission. The second and third components have the same absorption column density. This model was applied 
to the NW rim and the S rim, and the results are shown in Table 2. 

As expected from the work of Lu \& Aschenbach [29] there is a low temperature component 
({\it{vmekal1}}) with a temperature between 37 and 44 eV associated 
with a column density around 1.3$\times 10^{21}$ cm$^{-2}$, which is about a factor of two to three higher than found by Lu \& Aschenbach 
[29]. But the column density for the high temperature ({\it{vmekal2} + power law}) components is higher by another factor of three to four, which would indicate 
a larger distance if the spectral model is correct. Formally, i.e. based on ${\chi}^2_{\nu}$  (c.f. Table 2), the fits are acceptable for both 
the NW and S region, with no significant difference of the best fit parameters for the two regions. 
The power law slopes of 2.59 and 2.55, respectively, and the absorption column 
densities agree remarkably well with the {\it{ASCA}} measurements [44,48] 
despite the significantly higher photon statistics and higher energies covered by {\it{XMM-Newton}}.

Because of the ambiguity concerning the contribution of the Vela SNR the X-ray spectral analysis is usually restricted to the energy range E$>$0.8 keV.

\subsubsection{The north-western rim}

Figure 4 shows the {\it{XMM-Newton}} EPIC-PN image of the north-western rim. A bright filament like structure defines the outer edge of the 
remnant and a second, non-aligned and significantly fainter filament like structure is seen further inside. The two structures seem to join each other 
at the north-eastern tip. This X-ray image confirms to the image published by Iyudin et al. [22], which resolves the remnant's outer boundary into a number of fine filaments. The image also shows a number of point sources near the NW rim. 
\begin{figure}[hbt]
  \centering
\includegraphics[bb=0 0 451 283,width=8.0cm,clip]{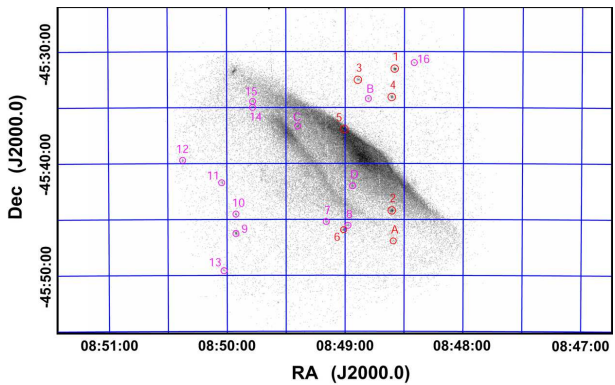}
\caption{The  deep exposure {\it{XMM-Newton}} EPIC-PN image shows the bright NW part of the RX J0852.0-4622 clearly resolved, and some new, previously unknown X-ray point-like sources. This EPIC-PN image was produced from EPIC-PN calibration data taken during {\it XMM-Newton} revolutions 0367, 0533, 0534, and 0897. }
\end{figure}

\begin{center}
\begin{table*}[bht]
\caption{Spectral fit results with a {\it wabs(vmekal)+wabs(vmekal+power law)} model for the range 0.2$\leq$E$_x$$\leq$10 keV.}

\begin{tabular*}{166mm}{||l||c|c|c|c|c|c|c||@{\extracolsep\fill}} \hline
& low kT$_1$ & high kT$_2$ & $N_H$(kT$_1$) &$N_H$(kT$_2$) & power law &  $\chi$$^2$ (dof) & ${\chi}^2_{\nu}$  \\

\raisebox{1.5ex}[0pt]{Region} & (eV) & (keV) &(10$^{22}$ cm$^{-2}$) &(10$^{22}$ cm$^{-2}$) & photon index&  & \\
\hline
\hline
NW rim  & 37.5$^{+4.0}_{-2.0}$ & 3.84$^{+1.65}_{-1.30}$  & 0.138$^{+0.022}_{-0.026}$ & 0.461$^{+0.023}_{-0.019}$ & 2.59$^{+0.16}_{-0.07}$ & 796.48 (777) & 1.025 \\
\hline
 Southern rim 1  & 43.5$\pm$0.8 & 2.92$\pm$2.50  & 0.123$\pm$0.016 & 0.471$\pm$0.035 & 2.55$\pm$0.25 & 349.01 (379) & 0.92 \\
\hline
 \hline
\end{tabular*}
\end{table*}
\end{center}       

\begin{center}
\begin{table*}[bht]
\caption{Fit results for E$_x$$\geq$0.8 keV with various spectral models for the continuum including a power law, the {\it sresc} model with and without 
one or two Gaussian lines.}

\begin{tabular*}{154mm}{||l||l||c|c|c|c|c||@{\extracolsep\fill}} \hline
 &  & ph. index, or &$N_H$ & radio index &  $\chi$$^2$ (dof) & ${{\chi}^2}_{\nu}$  \\

 \raisebox{1.5ex}[0pt]{Region} &  \raisebox{1.5ex}[0pt]{Model}& ${\nu}_{rolloff}$, Hz &(10$^{22}$ cm$^{-2}$) & $\alpha$ at 1 GHz&  & \\
\hline
\hline
 NW rim  & {\it powerlaw} & 2.60  & 0.496 & --- & 166.64 (173) & 0.963 \\
\hline
 NW rim  & {\it sresc} & (2.2$^{+0.4}_{-0.2}$)$\times$10$^{17}$  & 0.389 & 0.24 & 157.76 (172) &  0.917 \\
\hline
 NW rim  & {\it sresc+gauss} & 2.20$\times$10$^{17}$  & 0.389 & 0.24 & 147.06 (170) & 0.865 \\
\hline
 NW rim  & {\it sresc+2gauss} & 2.20$\times$10$^{17}$  & 0.389 & 0.24 & 145.04 (170) & 0.853 \\
\hline
 Southern rim  & {\it sresc} & (2.6$^{+0.6}_{-0.4}$)$\times$10$^{17}$  & 0.414 & 0.31 & 230.707 (225) & 1.025 \\
\hline
 \hline
\end{tabular*}
\end{table*}
\end{center}

In an attempt to improve upon the non-thermal component modelling of the spectrum with a straight power law model, we have 
tried in addition the
synchrotron model ({\it sresc}) of XSPEC, developed by Reynolds [40,41]. This model represents synchrotron
emission from the shock wave accelerated high-energy electrons and which takes care of  
electron escape or synchrotron losses 
by a steepening of the spectrum towards higher 
energies. 
As can be seen from Table 3 the reduced ${\chi}^2_{\nu}$ value is slightly lower than for a 
straight power law. Figure 5 (left) shows two local residuals around $\sim$4.4 keV and $\sim$6.5 keV that were found from the fit to the GTO data of rev. 0252.   
By adding one or two Gaussian  shaped line(s) {\it{gauss}} at these energies to the {\it sresc} continuum
model it is possible to further lower ${\chi}^2_{\nu}$ [22]. We note that a simple power law model is quite
acceptable judging on the total ${\chi}^2$ value. We also note that to cover the low energy part of spectrum we need a thermal component
which is represented by a {\it{vmekal}} model. 
\begin{figure*}[hbt]
  \centering
\includegraphics[bb=100 575 500 712,width=17.5cm,clip]{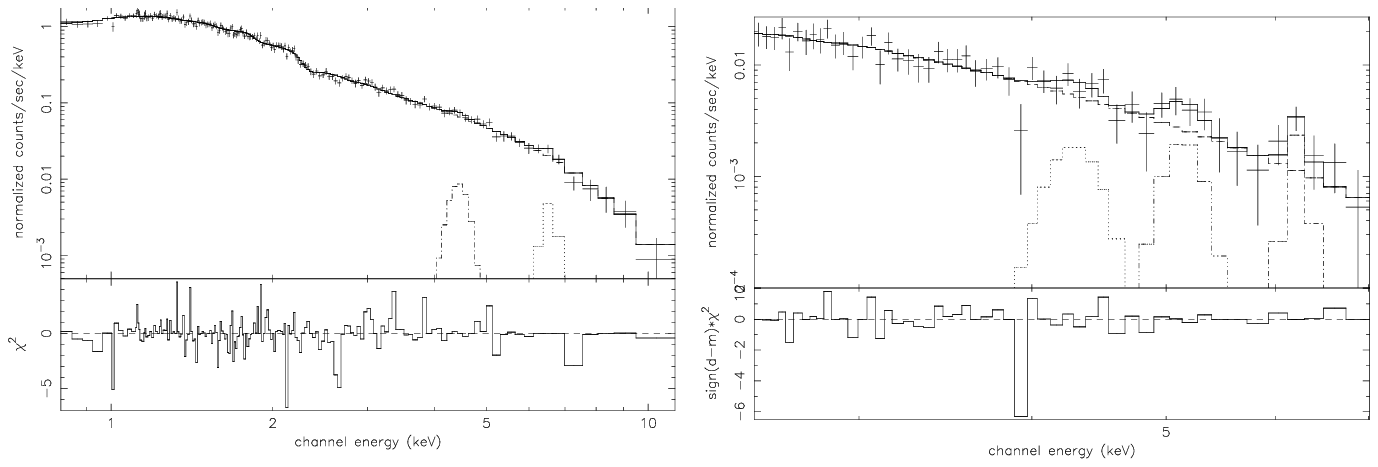}
  \caption{Left: PN spectrum of the brightest region of the RX J0852.0-4622 NW rim (region 1). Right: EPIC-PN
spectrum of fainter filament of the NW rim (region 2). The spectra were obtained from a combination of observations dedicated to EPIC-PN calibration (XMM-Newton revolutions 0367, 0533, 0534, and 0897). Spectra were fitted with the {\it sresc+Ngauss} model.}
\end{figure*}
We have also analysed the X-ray spectrum of each of the two filaments (c.f. Fig. 4) separately. 
Both filaments have a power law continuum spectrum
for E$_x \geq$0.8 keV, which can be also well fitted by the {\it{sresc}} model of XSPEC 
(c.f. Fig. 5). The line feature at $\sim$4.2 keV
detected by {\it{ASCA}} [48], is detected by {\it{XMM-Newton}} [22], and is present in both filaments.
The line flux is larger in the brighter (outer) filament, but the line is more clearly detected in the fainter filament. Additionaly, two more lines are clearly observed in the fainter filament, which has smaller continuum flux that the brighter filament. We believe that the X-ray line at $\sim$4.2 keV is a direct
consequence of the $^{44}$Ti decay in the SNR shell. Note the sharp outer boundary of the remnant and the
clear presence of at least two arc-like features. The fainter one might correspond to the reverse shock, that only starts to develop. One of the line-like excesses, detected in the spectra of NW rim,
is also visible in the spectra of
the southern and western rims (Fig. 6).
Namely, the strongest line at E$_{line}$=4.24$^{+0.18}_{-0.14}$ keV of the NW reg. 1 for a power law continuum model, and
E$_{line}$=4.44$\pm$0.11 keV for the same region using the {\it sresc} model, is also detected in the spectra of the southern and western rims.
\begin{figure*}[hbt]
  \centering
\includegraphics[bb=100 575 500 712,width=17.5cm,clip]{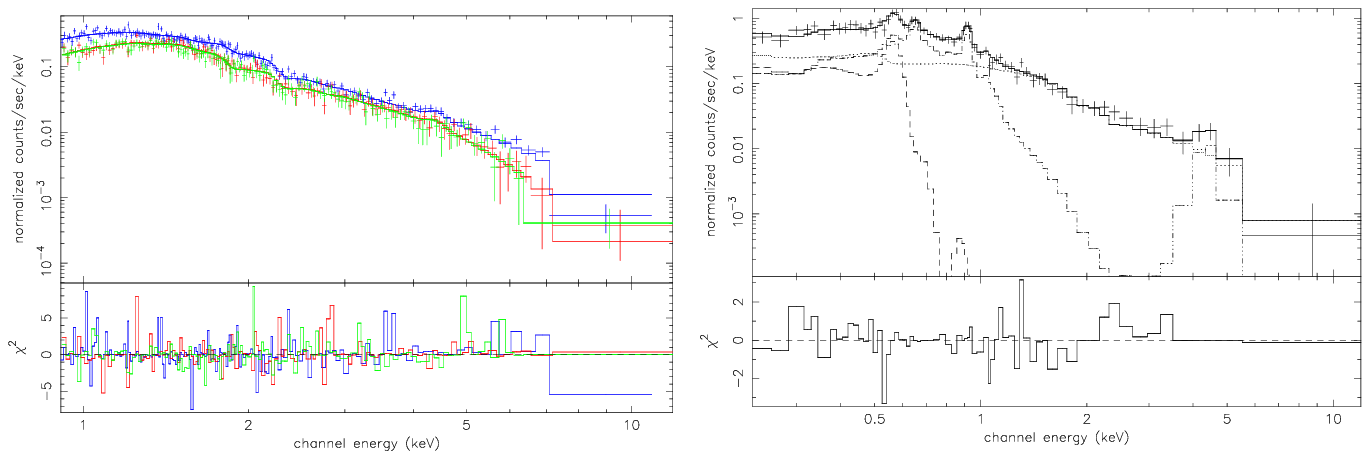}
  \caption{ Left: EPIC-PN and EPIC-MOS1,2 spectra of the brightest region 1 of the RX J0852.0-4622 southern rim 
for E$>$ 0.8 keV. Spectra were fitted with a {\it vmekal+sresc+gauss} model. Right: MOS spectrum of the western rim
for E$>$ 0.2 keV, in comparison.
Spectrum was fitted with a {\it 2vmekal+powerlaw+gauss} model.}
\end{figure*}

\par
We confirm, that the emission line feature at $\sim$4.2 keV found by {\it{ASCA}} in the NW rim spectrum of RX J0852.0-4622 is also found 
in the {\it{XMM-Newton}} data (Fig. 5, 6). Our values of the line energy are slightly higher than the value given by Tsunemi
et al. [48], but are consistent with the {\it{ASCA}} SIS0 spectrum of the NW rim (Fig. 4 in the paper [48]).
The complete data set of lines is consistent with one value 
for the line energy common to all three 
sections of the rim, which is  
$E_x = 4.45 \pm 0.05$ \hspace{2mm}keV. We believe that the X-ray line at $\sim$4.2 keV is a direct consequence of the $^{44}$Ti decay in the SNR shell.
\par
The significance values of the line detection depend on the region from which the background model was formed, and
not as yet conclusive to claim the existence of the line beyond any doubt, although the coincidence of the line
energies is rather compelling.
   
\subsubsection{The radial profile of the NW rim}

The new, combined EPIC-PN image of the northern rim of RX J0852.0-4622 is shown in Figure 7. A structure of rectangular boxes is overlaid. For each individual box radial profiles across the rim were constructed.
The radial profiles clearly show the presence of the SNR forward shock and of the juvenile reverse shock. The profile shown on the very right side of Fig. 7 is likely containing two filaments that may constitute the forward shock. It will be interesting to follow up this profile structure with better statstics and angular resolution.
\par
{\it Chandra} also observed this part of the NW rim [4,33].
By fitting the radial profiles with a suitable function Bamba et al. [4] derived the magnetic field in the forward shock. The magnetic field and the angular size of the forward shock were used by Bamba et al. [4] to constrain the distance and the age of the Vela Jr. SNR. Values cover the range from 420 yr to 1400 yr for the SNR age, and from 260 pc to 500 pc for the SNR distance.
\par
These estimates of the age and of the distance to SNR are not stronly dependent on the ambient number density or the ejecta mass, and can therefore accomodate ejecta masses of both type Ia and core-collapse SNe.
Although the allowed parameter regions are too broad to derive strong conclusions, it is possible to state that indeed Vela Jr. is a nearby and young SNR.

\begin{figure*}[hbt]
  \centering
\includegraphics[bb=70 590 526 712,width=17.5cm,clip]{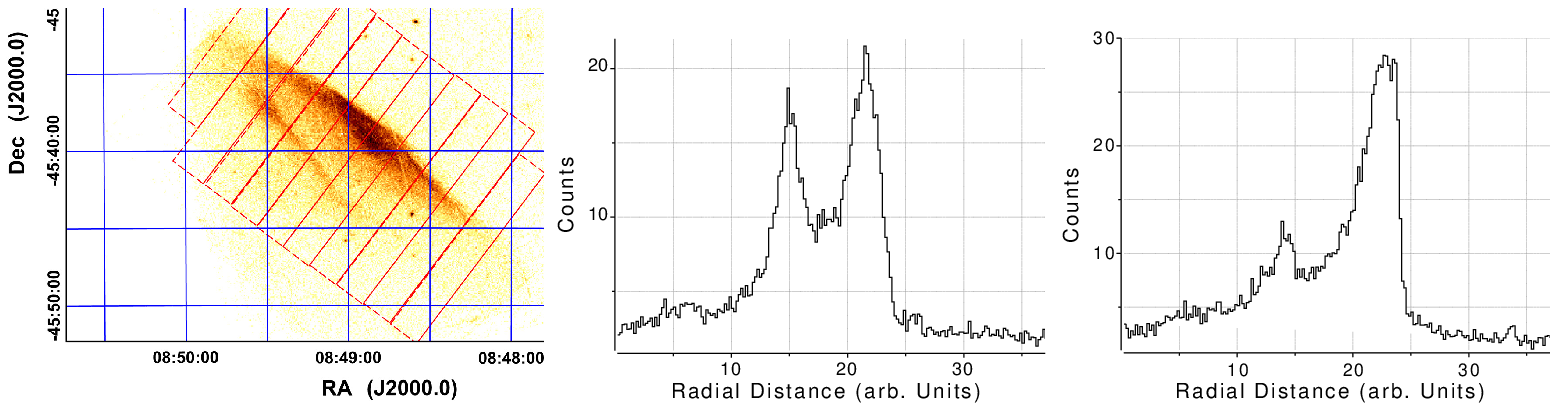}
\caption{ Overlay of the X-ray image and a grid of rectangular boxes used to build radial profiles of the SNR NW rim (left). Examples are shown for box \#3 (middle of figure) and box \#5(right); box counting is from left to right. The profile of box \#5 profile shows fine filaments in the radial profiles of the SNR NW rim.}
\end{figure*}
\subsection{The western rim}

While the NW rim spectra are quite distinct from the underlying thermal emission of the Vela SNR that is dominating
at E$_x$$\leq$0.8 keV, the western rim spectra for E$_x$$\geq$0.8 keV show apart from a steep power law component
with a photon index of $\sim$4.7 
an additional 
thermal component, which is characterized by N$_H$$\sim$(2.84$\pm$0.15)$\times$10$^{21}$ cm$\sp{-2}$ and kT$\sim$0.2 keV.
Both the power law and the thermal component are significantly different from those of the NW and S rim. 
We note that the PN observation of the W rim was heavily contaminated by  solar proton flare events that 
lowered the 
effective exposure by a factor of three when the {\it{good time interval}} selection process 
was applied. The MOS spectra were
not strongly affected by the soft protons and confirm the line emission at 4.4$\pm$0.18 keV, also for the western rim.
\par
The western limb shows up  in the X-ray image of ROSAT as the X-ray bright feature W (Fig. 1).
We note that the radio emission from this part of RX J0852.0-4622 at 2420 MHz is rather weak.
This low  radio-continuum emission is consistent with the fit to the X-ray spectrum  (Fig. 6, right), which reveals 
a relatively small 
contribution of the synchrotron radiation in this region of the SNR.

\section{Gamma-ray observations of Vela Jr.}
\subsection{INTEGRAL results and expectations}
There have been numerous discussions about the significance of the 1.157 MeV line detection from this SNR (see e.g.
[3,21,22,43]) and it is obvious that an independent confirmation of the detection is needed.
First attempts made with INTEGRAL have so far produced an upper limit, which is about three times as high as the
flux measured with {\it{COMPTEL}} [27]. Moreover, if estimates of the Vela Jr. ejecta velocity made by COMPTEL is indeed about 15.000,0 km/s [19], the chances to detect the 1.157 MeV line are fairly low due to the very large Doppler width of the line. The chance to detect line emission from GRO J0852-4642 with SPI is reasonably high only for the line of excited $^{44}$Sc at 78 keV. In this specific case the detection at $\sim$3 $\sigma$ significance is expected for good quality exposures of $\ge$7$\times$10$^6$ seconds. Hopefully this integrated exposure time will be reached in 2007.
\par
The X-ray emission line feature at $\sim$4.2 keV found by {\it{ASCA}} [48] in the NW rim spectrum of RX J0852.0-4622 was confirmed 
by {\it{XMM-Newton}} data (Fig. 5, and [22]), and by {\it Chandra} [4]. The XMM-Newton
spectra of the SNR western and southern rims show the line as well (Fig. 6), at an energy which is consistent
with that of the NW rim. 
\par
The complete set of X-ray lines detected by ASCA, {\it Chandra} and XMM-Newton is consistent with one value 
for the line energy common to all three 
sections of the rim, which is  
$E_x = 4.45 \pm 0.05$ \hspace{2mm}keV.
The detections of X-ray line from all X-ray bright rims of GRO J0852-4642 is pointing to a more or less homogeneous distribution of the line emitter along the SNR forward shock rims. 
\par
Quite unfortunately, this spatial distribution of the line emission over the SNR makes the detection of the 68 keV and 78 keV lines of $^{44}$Sc from GRO J0852-4642 by IBIS-ISGRI quite difficult. To detect the $^{44}$Sc lines IBIS-ISGRI needs to accumulate an exposure equivalent of a few 10$^8$ seconds in the most optimistic scenario of the SN explosion, e.g. if the $^{44}$Ti is contained in a few well defined clumps of the SNR ejecta. If the $^{44}$Ti mass is rather smoothly distributed (diluted) along the outer rim of an expanding ejecta, and this is the case that is indicated by the X-ray line detections, the detection of the Sc lines at 68 keV and 78 keV will even need an order of magnitude higher exposure, which is out of reach. To estimate the exposure time of ISGRI one is advised to have a look at the paper by Renaud et al. [39] where the extended sources detection by ISGRI is critically evaluated.  
\par
Note that the situation is completely different for the Cas A SNR where  $^{44}$Sc distribution is consistent with being a point-like excess for IBIS-ISGRI. This explains a relatively easy detection of two $^{44}$Sc lines from this SNR by IBIS-ISGRI [40,51], while the high velocity of the ejecta containing $^{44}$Ti ($\sim$8000 km/s) makes a detection  by SPI of the 78 keV line from Cas A quite a demanding excercise [9,51].

\subsection{VHE $\gamma$-ray observations}
Recently H.E.S.S. published results of the first of likely many more observations of this shell-type SNR [1]. The spectrum derived by H.E.S.S. from RX J0852.0-4622 in the energy range of 0.5 TeV to 8.0 TeV is a power-law with a photon index of $\sim$2.1. 
The earlier observation by CANGAROO-II of the NW rim of RX J0852.0-4622 and early results [24] are now revised after much longer observation of the whole SNR by CANGAROO-III and improved data analysis [15]. The latest spectral shape of RX J0852.0-4622 derived by CANGAROO-III in the energy range from $\sim$1.0 TeV to $\sim$5.0 TeV is consistent with that of H.E.S.S..
\par
The multiwavelength spectrum of the SNR involving the radio, X-ray and VHE $\gamma$-ray observations still does not really constrain the origin of the high-energy emission of this SNR. Both leptonic, and proton models of the high-energy emission are allowed by fits to the multi-$\lambda$ spectrum [1].

\section{GRO J0852-4642 spectrum and progenitor}
The spectral shape measured with {\it{XMM-Newton}} from RX J0852.0-4622 is basically the same as what has been measured with {\it{ASCA}}. 
The {\it{ASCA}} and recent {\it Chandra}  measurements indicate the presence of an emission line like feature at around 4.1 - 4.2 keV. 
The {\it{XMM-Newton}} data confirm such a feature, and it seems to be present everywhere on the remnant's rim. The line energy 
averaged over the three observational fields is 4.45 $\pm$ 0.05 keV. We attribute this line or lines to the emission of Ti and Sc which might be 
excited by atom/ion or ion/ion collisions. 
The X-ray line flux expected from such an interaction is consistent with the 1.157 MeV $\gamma$-ray line flux measured 
by {\it{COMPTEL}}. This consistency of the X-ray line flux and the $\gamma$-ray line flux 
lends further support to the existence and amounts of Ti in RX J0852.0-4622 claimed by Iyudin et al. [19] and to the suggestion 
that RX J0852.0-4622 
is young and nearby [3]. 
Iyudin et al. [19] quote a very large broadening of the 1.157 MeV $\gamma$-ray line which would indicate a large 
velocity of the emitting matter of about 15.000 km/s. Previously we believed that such a high ejecta velocity for Ti is found only in explosion models of 
sub-Chandrasekhar type Ia supernovae [28,53]. More recent calculations of the core-collapse SNe models with jet-like ejecta from the pole regions show high-velocity components of around 15.000,0 km/s, which contain $^{44}$Ti (Nomoto et al. [32], and references therein). Moreover, in recent observations of ``ordinary'' type Ia SNe high-velocity features in the early spectra of SNe were discovered [47]. 
Thus the question on the nature of the SN that has given a birth to RX J0852.0-4622 remains unanswered. The puzzle is waiting for to be solved.

\section*{Acknowledgments}

Authors acknowledge comments of the anonymous referee that helped to improve the quality of the paper. 
The XMM-Newton project is an ESA Science Mission with instruments and contributions directly funded by ESA Member States and the USA (NASA). The XMM-Newton project is supported by the Bundesministerium f\"ur Bildung und Forschung / Deutsches Zentrum f\"ur Luft- und Raumfahrt (BMBF / DLR), the Max-Planck-Gesellscahft and the Heidenhain-Stiftung.

\end{document}